\documentclass[jcp,aip,
11pt,
amsmath,amssymb,
floatfix,
reprint,
]{revtex4-1}

\usepackage[utf8]{inputenc} 
\usepackage[T1]{fontenc} 
\usepackage{verbatim}

\usepackage[table]{xcolor} 
\definecolor{CBred}{RGB}{215,25,28}
\definecolor{CBorange}{RGB}{253,174,97}
\definecolor{CByellow}{RGB}{255,255,191}
\definecolor{CBgreen}{RGB}{171,211,164}
\definecolor{CBlgreen}{RGB}{166,217,106}
\definecolor{CBdgreen}{RGB}{26,150,65}
\definecolor{CBblue}{RGB}{43,131,186}
\definecolor{CBblue2}{RGB}{146,197,222}
\definecolor{CBdblue}{RGB}{5,113,176}
\definecolor{CBgray60}{RGB}{102,102,102}
\definecolor{CBgray20}{RGB}{204,204,204} %
\definecolor{CBgray90}{RGB}{80,80,80} %
\definecolor{CBcyanlight}{RGB}{1,102,94}

\usepackage[
    pdftex,%
    unicode, 
    colorlinks=true,
    ocgcolorlinks=false, %
    linkcolor=CBcyanlight, %
    anchorcolor=black,%
    citecolor=CBcyanlight, %
    filecolor=magenta, 
    menucolor=red,
    urlcolor=CBblue, 
    linkbordercolor=red,
    citebordercolor=blue,
    filebordercolor=magenta,
    menubordercolor=red,
    urlbordercolor=cyan,
    plainpages=false, %
    pdfpagelabels, %
    hypertexnames=true, %
    breaklinks=true, 
    linktocpage %
]{hyperref}

\usepackage{graphicx}

\usepackage[caption=false]{subfig}

\usepackage{dcolumn}%
\usepackage{bm}%

\usepackage{listings}

\usepackage[version=3]{mhchem}
\usepackage{units}
\usepackage{siunitx}

\usepackage{xspace}
\usepackage[textsize=small,textwidth=1.5cm]{todonotes}
\graphicspath{{pics/}{pics/decay/}}

\newcommand{\Eqs}{Eqs.\xspace}

\newcommand{\eg}{\mbox{e.\,g.}\xspace}

\newcommand{\lit}[1]{Ref.~\citenum{#1}\xspace}
\newcommand{\lits}[1]{Refs.~\citenum{#1}\xspace}

\newcommand{\erw}[1]{\ensuremath {\langle{#1}\rangle}}

\newcommand{\braket}[2]{\ensuremath{ \langle #1 | \, #2  \rangle }}
\newcommand{\ketbra}[2]{\ensuremath{  | {#1} \rangle %
\langle {#2} |}}
\newcommand{\ket}[1]{\ensuremath{  | {#1} \rangle}}
\newcommand{\bra}[1]{\ensuremath{\langle {#1} | }}
\newcommand{\matrixe}[3]{\ensuremath{ \langle{#1} | \vphantom
        {#1 #3} {#2}
| {#3} \rangle }}
\newcommand{\Matrixe}[3]{\ensuremath{ \left \langle{#1} \left| \vphantom {#1 #3} {#2}  \right| {#3} \right \rangle }}
\newcommand{\totd}[2]{\ensuremath{ \frac{\dd {#1}} {\dd {#2}} }}

\newcommand{\totdd}[2]{\ensuremath{ \frac{\dd^2 {#1}}{\dd {#2}^2} }}

\newcommand{\partd}[2]{\ensuremath{ \frac{\partial {#1}}
{\partial {#2}} }}

\newcommand{\partdd}[2]{\ensuremath{ \frac{\partial^2 {#1}}
{\partial {#2}^2} }}

\newcommand{\ii}{\ensuremath{\mathrm{i}}}

\newcommand{\dd}{\ensuremath{\mathrm{d}}}

\definecolor{ocre}{RGB}{243,102,25}
\definecolor{mygray}{RGB}{243,243,244}
\definecolor{fzjred}{RGB}{175,90,80}

\definecolor{blau}{HTML}{1F78B4}
\definecolor{gruen}{HTML}{33A02C}
\definecolor{hellblau}{HTML}{A6CEE3}
\definecolor{hellgruen}{HTML}{B2DF8A}

\definecolor{nrot}{HTML}{d7191c}
\definecolor{norange}{RGB}{253,174,97}
\definecolor{ngruen}{HTML}{abdda4}
\definecolor{nblau}{HTML}{2b83ba}

\definecolor{nrot1}{RGB}{215,48,31}
\definecolor{nrot2}{RGB}{252,141,89}
\definecolor{nrot3}{RGB}{253,204,138}
\definecolor{nrot4}{RGB}{254,240,217}

\definecolor{CBred}{RGB}{215,25,28}
\definecolor{CBorange}{RGB}{253,174,97}
\definecolor{CByellow}{RGB}{255,255,191}
\definecolor{CBgreen}{RGB}{171,211,164}
\definecolor{CBlgreen}{RGB}{166,217,106}
\definecolor{CBdgreen}{RGB}{26,150,65}
\definecolor{CBblue}{RGB}{43,131,186}
\definecolor{CBblue2}{RGB}{146,197,222}
\definecolor{CBdblue}{RGB}{5,113,176}
\definecolor{CBgray60}{RGB}{102,102,102}
\definecolor{CBgray20}{RGB}{204,204,204}

\newcommand{\new}{}
\let\oldtheequation\theequation
\makeatletter
\def\tagform@#1{\maketag@@@{\ignorespaces#1\unskip\@@italiccorr}}
\renewcommand{\theequation}{(\oldtheequation)}
\makeatother 

\begin{document}
  \title{Control of concerted back-to-back double ionization dynamics in helium}
\author{Henrik R.~Larsson}
\email{larsson [at] caltech.e$\delta$u}
\altaffiliation[Present address: ]{Division of Chemistry and Chemical Engineering, California Institute of Technology, Pasadena, CA 91125, USA}
\affiliation{Institut für Physikalische Chemie, Christian-Albrechts-Universität zu Kiel, Olshausenstraße 40, 24098 Kiel, Germany}
\affiliation{Department of Chemical Physics, Weizmann Institute of Science, 76100 Rehovot, Israel}
\author{David J.~Tannor}
\affiliation{Department of Chemical Physics, Weizmann Institute of Science, 76100 Rehovot, Israel}
\date{\today}
\begin{abstract}
Double ionization (DI) is a fundamental process that despite its apparent simplicity provides rich opportunities for probing and controlling the electronic motion. Even for the simplest  multielectron atom, helium, new DI mechanisms are still being found.
To first order in the field strength, a strong, external field doubly-ionizes the electrons in helium such that they are ejected into the same direction (front-to-back motion).
The ejection into opposite directions (back-to-back motion) cannot be described to first order, making it a challenging target for control.
Here, we address this challenge and optimize the field with the objective of back-to-back double ionization, using a (1+1)-dimensional model.
The optimization is performed using four different control procedures: (1) short-time control, (2) derivative-free optimization of basis expansions of the field, (3) the Krotov method and (4) control of the classical equations of motion. 
All four procedures lead to fields with dominant back-to-back motion.
All the fields obtained exploit essentially the same two-step mechanism leading to back-to-back motion:
first, the electrons are displaced by the field into the \emph{same} direction. 
Second, after the field turns off, the nuclear attraction and the electron-electron repulsion combine to generate the final motion into \emph{opposite} directions for each electron. 
By performing quasi-classical calculations, we confirm that this mechanism is essentially classical.
\end{abstract}

\maketitle
\section{Introduction}
\label{sec:intro}

Helium, despite its simplicity, can show rich, correlated dynamics of its electrons.\cite{el_el_corr_rev_liu_2011,staudte_2007,HHG_He_dudovich_2017,ueda_roadmap_2019,amini_symphony_2019}
One interesting process that appears in strong laser fields is double ionization (DI).\cite{multiple_ioniz_rev_dorner_2002,many_el_phys_rev_becker_2008,el_el_corr_rev_liu_2011}
Besides its fundamental importance for understanding light-matter interactions, double ionization provides new challenges for resolving and controlling attosecond-timescale dynamics of atoms and molecules.\cite{winney_attosecond_2017,li_multiorbital_2019}
DI can be generated by three main types of mechanisms, 
direct, sequential, and nonsequential double ionization.
In direct ionization, the energy of one photon is sufficient to ionize both electrons 
directly. %
In sequential ionization, one photon ionizes the first electron and a second photon ionizes the remaining electron. 
This requires a high photon energy 
(larger than the
second ionization potential, $I_2=\unit[54.4]{eV}$ for helium~\cite{he_NIST_ASD}) 
or an intensity large enough for over-the-barrier ionization. %
When the energy of the photon is smaller than the
second ionization potential, nonsequential double ionization can occur.  This requires an interaction with several photons quasi-simultaneously. 
The simultaneous motion of the electrons in nonsequential double ionization provides a rich physics with many different mechanisms in various regimes of 
\new{field intensities and energies}, ranging from infrared (IR) to extreme ultraviolet (XUV).\cite{el_el_corr_rev_liu_2011,many_el_phys_rev_becker_2008,multiple_ioniz_rev_dorner_2002}

One central feature of double ionization
is the final relative motion of the electrons. 
\new{For a linearly polarized field,} 
are they accelerated to the same or to opposite
directions?\cite{zhang_2014c,maxwell_2016,rudenko_recoil-ion_2008,staudte_2007,moshammer_few-photon_2007,EUVphotoninduced2009jiang,katsoulis_slingshot_2018,Nonsequential2014kubel,Correlated1994becker,Double2010chen,Laserassisted2014liu,Strongfieldinduced2020borisova}
The motion into the same direction is called front-to-back or correlated motion whereas the motion into opposite directions is called back-to-back or anti-correlated motion.\cite{staudte_2007,zhang_2014c,bergues_2012,ngokodjiokap_2014,liu_strong-field_2008,rudenko_recoil-ion_2008}
The external field couples directly only to the front-to-back motion. Hence, to first order, the field can accelerate both electrons simultaneously only in the same direction.
In many situations, front-to-back motion is the dominant contribution to double ionization.\cite{prauzner-bechcicki_2007,lein_2000,Laserinduced2003eremina,bergues_2012}
However, in certain regimes such as the two-photon nonsequential double ionization regime (photon energy above $\unit[39.5]{eV}$) back-to-back motion is dominant.\cite{moshammer_few-photon_2007,
rudenko_recoil-ion_2008,EUVphotoninduced2009jiang,Decoding2008horner,Evidence2008foumouo}
Hence, the interplay between the electron-electron, electron-nuclear and electron-field interactions are nontrivial and lead to many different types of mechanisms and effects,
including surprising transitions from front-to-back to back-to-back motion, depending on the field parameters.\cite{el_el_corr_rev_liu_2011,Nonsequential2014kubel,katsoulis_slingshot_2018}

The rich physics of nonsequential double ionization even in the simplest system of the helium atom makes this process interesting and challenging for quantum control optimization.  
One of the central aims of quantum control theory is to
maximize observables of a wavefunction by steering it from an initial state via a time-dependent perturbation to a final state.
Typically, the perturbation is an external electromagnetic field interacting with the dipole moment and the initial state is the groundstate. The field is modified until the observable at the final time reaches an optimum. 
By using different types of algorithms for optimizing the field and by constraining the field to certain shapes or intensities, new insights can be gained and new physical mechanisms can be discovered.%
\cite{oct_tannor_1985,oct_kosloff_1989,cooling1999tannor,crab2011doria,oct2016goetz,RemoteOct2018heck,oct2020larrouy,Optimal2005papastathopoulos,Quantum2002sukharev}
While there is an extensive amount of work on controlling laser-matter interactions via selected field parameters such as carrier envelope phase or polarization, 
so far the application of full control theory 
to strong-field science and to double ionization in particular has been limited. 
Some selected applications to date include control of single ionization,\cite{oct2016goetz,elIoniz2016goetz,hhg2016schonborn,hhg2020schaefer}
and experimental pulse optimization for double ionization with IR pulses.\cite{Optimal2005papastathopoulos}

The objective of this article is 
twofold. The first objective is to introduce optimal control algorithms to double ionization dynamics and to compare different control procedures.
The second objective, as an initial example, is to control back-to-back double ionization by an external field.
As we will show, due to the necessity of higher-order effects and many-body contributions, this objective is difficult to achieve and requires a careful choice of the objective to be optimized.
In passing, we note that this objectives bears a resemblance to controlling the three-body dissociation of an aligned triatomic molecule along the symmetric or anti-symmetric stretch motion.  

We use four different control procedures: (1) short-time  control, where the control objective is maximized at each time step, (2) derivative-free optimization using a basis representation of the field, (3) the Krotov algorithm and (4) control of the classical equations of motions.
We show that a fifth procedure, local control, is not appropriate for our objective but its analysis helps in understanding the interplay between the field and the objective.

All four algorithms used have different strengths and weaknesses and lead to different optimized fields. 
However, we show below that for this system, all obtained fields lead to the same mechanism: a two-step procedure, where the initial wavepacket is first displaced by the field in the front-to-back direction and then propagated field-free toward the desired back-to-back direction. This mechanism is confirmed by quasi-classical calculations.
Thus, the essentials of the mechanism are classical.

Optimal control algorithms typically require tens or
hundreds of wavepacket propagations with external fields. This and the computational demands of double ionization in general (a two-particle continuum needs to be adequately described) limits us here to use a (1+1)-dimensional model of the helium atom. This reduced dimensional model can, however, capture the main physics of the full (3+3)-dimensional system.\cite{lein_2000,prauzner-bechcicki_2007,Double2010chen,tdgasci_LiH_bonitz_2016,tdgasci_1D_bonitz_2016,yu_2016b}
Although we do not predetermine
particular pulse shapes and intensities for the control, 
\new{to reduce overall computational cost} 
we restrict the field duration to a maximal $\unit[1.45]{fs}$, which clearly limits the fields to be in the XUV (and X-ray) and few-cycle IR regimes.\cite{Multiphoton2017pindzola,Laserassisted2014liu}
Aside from these restrictions, the field can \new{take} any possible shape and intensity required to reach the objective. 
While the fields obtained by such a procedure may not be directly experimentally accessible, we extract the mechanism by which the field leads to the desired objective. 
The key characteristics of the field, e.g.~asymmetry of the pulse and intensity patterns, should then be able to be implemented experimentally. \new{The required intensities, pulse durations, and pulse energies we found should be within reach of current Free Electron Lasers.\cite{moshammer_few-photon_2007,rudenko_recoil-ion_2008,Highintensity2015marinelli,Terawatt2021emma,Using2021bergmanna}}

Note that although the model is of reduced dimensionality, it is extremely challenging computationally. The many long propagations required for the control methods used here were possible because of an efficient new implementation of the wavepacket dynamics.\cite{pW_tannor_2016,DCO_hartke_2018,pvb_rev_tannor_2018,dpmctdh_tannor_2017}

The outline is as follows. \autoref{sec:model_system} introduces the model system and the coordinates used.
\autoref{sec:target}
defines the regions corresponding to mutually exclusive ionization outcomes and introduces the objectives to be optimized.
The various control methods used are presented in \autoref{sec:control_methods}.
The results are presented in \autoref{sec:results} and discussed in \autoref{sec:discussion}. We conclude in \autoref{sec:conclusions}.

\section{Model system}
\label{sec:model_system}

In the following, we use the standard regularized ($1+1$)-dimensional model of helium, where the quantum mechanical Hamiltonian takes the form of~\footnote{Throughout, we use atomic units ($m_e = e = 4\pi\epsilon_0=a_0=2|E_\text{Ryd}|=1$) unless indicated otherwise.}
\begin{align}
 \hat H &= \hat H_0  + \hat X A(t),\\
 \hat H_0 &= -\frac{1}{2} \partdd{}{x_1} - \frac12 \partdd{}{x_2} + V(x_1,x_2),
 \label{eq:he_hamilt_a}\\
 V(x_1,x_2) &= \sum_{i\in\{1,2\}} -\frac{2}{\sqrt{x_i^2 + \alpha^2}} + %
  \frac{1}{\sqrt{(x_1-x_2)^2 + \alpha^2}},\\
  \hat X &= -\ii \left( \partd{}{x_1} + \partd{}{x_2}  \right), \label{eq:he_x_a}
\end{align}
where $x_i$ is the coordinate of electron $i$. 
The regularization parameter $\alpha$ is taken from \lit{pvb_edyn_tannor_2015} and set to  $0.739707902$.
This regularization leads to a match of the exact ground state energy of the helium atom.
The interaction with the external field is described by the vector potential $A(t)$ and the interaction operator $\hat X$ in {(reduced)} velocity gauge.
{An additional term, $A^2(t)/2$, in the Hamiltonian has been eliminated by a standard gauge transformation,\cite{bransden_joachain_book}
which does not affect the expectation values of interest discussed here.}
We prefer velocity gauge over length gauge because only fields $E(t)=-\dd A(t)/\dd t$ with vanishing time integral are physical.\cite{madsen_gauge_2002}
This can easily be achieved by requiring that the potential $A(t)$ vanishes for $t\to \pm\infty$. Additionally, the velocity gauge typically is computationally more efficient, especially at higher pulse intensity,\cite{madsen_2010}
and the local control expressions shown in \autoref{sec:theory_lct} are simpler in velocity gauge.

To take the \new{fermionic} nature of the electrons into account, we write the total wavefunction as direct product of a spin-dependent and a coordinate-dependent part. Since the non-relativistic Hamiltonian does not couple different spins, the spin-dependent part of the wavefunction remains constant during the time evolution.
For helium, the electronic ground state is a singlet state whose spin-dependent part is antisymmetric with respect to exchange of the electrons, $2^{-1/2} [ \ket{\alpha_1\beta_2} - \ket{\beta_2\alpha_1} ]$. 
Thus, to ensure total antisymmetry of the wavefunction with respect to exchange of the electrons, the \emph{coordinate}-dependent part of the wavefunction has to be \emph{symmetric}, $\psi(x_1,x_2) \stackrel!= \psi(x_2,x_1)$.
Symmetry can be exploited by rotating the coordinate system by 
$45^\circ$. In this rotated coordinate system, the coordinates $u$ and $v$ are defined as
\begin{align}
u &= \frac{x_1+x_2}2,&  v &= \frac{x_1-x_2}2,\label{eq:u_v_sys}\\
\Leftrightarrow u+v &= x_1,& u-v &= x_2.\label{eq:u_v_sys_inv}
\end{align}
Exchange of $x_1$ and $x_2$ leads to a sign flip in $v$. Accordingly, the wavefunction is 
axially symmetric in $v$,
\begin{equation}
\psi(u,v) = \psi(u,-v).
\end{equation}
We will exploit this symmetry by using only symmetric basis functions in the $v$ coordinate, thus decreasing the number of required basis functions by two, compared to the unrotated coordinate system; see \new{Section A} in the supplementary material. %

\begin{figure}
    \includegraphics[width=\columnwidth]{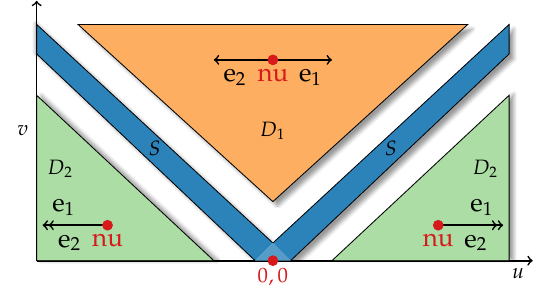}
  \caption{
  Regions of the one-dimensional helium atom in the continuum for the rotated coordinate 
system [\autoref{eq:u_v_sys}]. The red dot marks the origin where the ground-state density is peaked. The blue 
region $S$ corresponds to single ionization and the two regions $D_{\{1,2\}}$ denote 
double ionization and are not connected. 
$D_1$ ($D_2$) corresponds to double ionization where the electrons are ionized into 
opposite (same) directions. The motions of the two electrons are depicted by the small diagrams; showing the nucleus as a red dot. %
The white region in between denotes an 
intermediate between single and double ionization. 
The regions are only shown for values of $v$ larger or equal to zero. For 
negative values, the schematic is mirrored; compare with \autoref{fig:symCoulPot}. 
}
  \label{fig:fig_he_regions}
\end{figure}

Besides exploiting symmetry, this coordinate system allows for a very simple 
interpretation of the 
wavefunction in terms of the relative motion of the electrons; see 
\autoref{fig:fig_he_regions}.
Comparing the electronic motion in $u$ and $v$ with the vibrational normal modes of a diatomic system, $v$ describes a symmetric stretch mode  %
and $u$ describes an antisymmetric 
stretch %
mode. Large values of $v$ (compared to $u$) indicate large separations of the two electrons: 
They are on different sides of the nucleus. 
In contrast, large values of $u$ (compared to $v$) indicate that they are on the same side
of the nucleus and that they approach each other with decreasing $v$.

\new{Inserting Eqs.~\ref{eq:u_v_sys_inv} into Eqs.~\ref{eq:he_hamilt_a} and \ref{eq:he_x_a} leads to the definitions of the}
field-free Hamiltonian and the field-interaction operator in this rotated coordinate 
system: %
\begin{align}
 \hat H_0 =& -\frac14 \left(\partdd{}{u} + \partdd{}v\right) + V(u,v),\label{eq:he_H_s}\\
 \begin{split}
V(u,v) =& %
\frac1{\sqrt{(2v)^2+\alpha^2}} - \frac2{\sqrt{(u+v)^2+\alpha^2}} \\ &-  
\frac2{\sqrt{(u-v)^2+\alpha^2}},
\end{split}\label{eq:he_V_s}\\
\hat X =& -\ii \partd{}{u}. \label{eq:he_B_s_v}
\end{align}

\begin{figure}
    \includegraphics[width=\columnwidth]{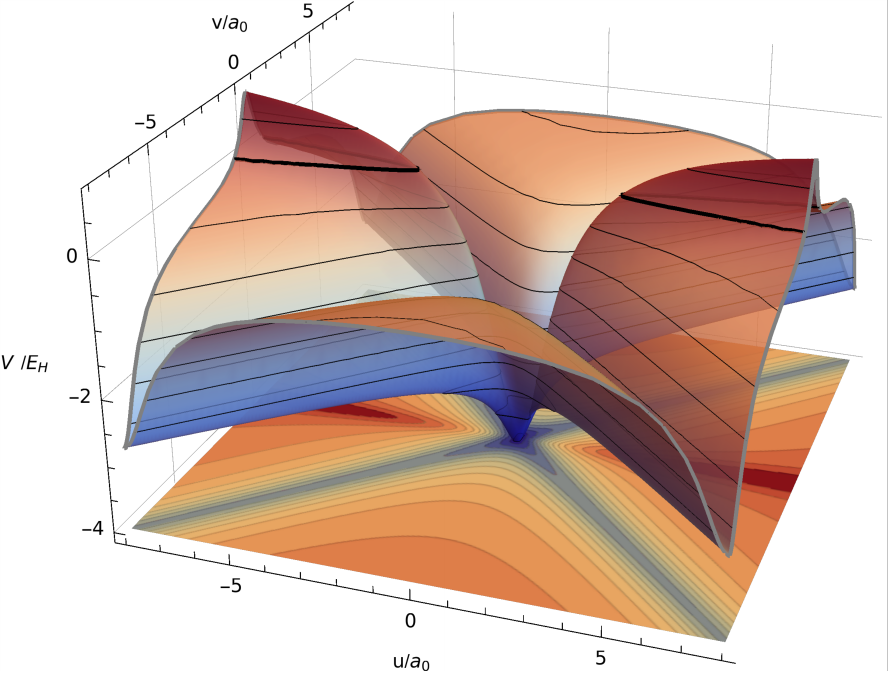}
  \caption{Potential $V(u,v)$ [\autoref{eq:he_V_s}] of the helium model in 
the rotated coordinate system [\autoref{eq:u_v_sys}]. The contour line at $V(u,v) = 0$ is highlighted by thicker lines, and parts of the potential are plotted transparently in order to recognize the part at $v>0$. Note the \new{axial and mirror} symmetry of the potential, $V(u,v) = V(u,-v) = V(-u,v) = V(-u,-v)$.}
  \label{fig:symCoulPot}
\end{figure}

The potential $V(u,v)$ is shown in \autoref{fig:symCoulPot}. 
Note that the repulsive electron-electron interaction is peaked along the $v=0$ line and 
the attractive electron-nuclear interaction is peaked along the diagonal $u=v$ lines. 

\section{Regions of Ionization}
\label{sec:target}
In the continuum, there are three regions with different characteristics. 
They are shown schematically in \autoref{fig:fig_he_regions}.
$S$ corresponds to single ionization. 
$D_1$ ($D_2$) corresponds to double ionization where the electrons are ionized into 
different (same) directions.
When double ionization to either $D_1$ or $D_2$ occurs, typically 
parts of the wavefunction remain in the single ionization region $S$.

The field interaction operator $\hat X$ in the rotated coordinate system, \autoref{eq:he_B_s_v}, reveals that an external field couples only with the $u$ direction. That is, both electrons are
accelerated into the same direction.
If the perturbation induced by the external field is stronger than 
the Coulomb repulsion of the electrons, they mainly occupy region $D_2$, indicating a correlated, or front-to-back process of the double ionization. 
An anticorrelated, back-to-back process is apparent if the electrons occupy region $D_1$.
Typically,\cite{prauzner-bechcicki_2007,lein_2000} 
both regions $D_2$ and $D_1$ are occupied but, due to the form of the perturbation induced by the 
field, there is 
a substantially larger fraction of the wavepacket in region $D_2$.%
\footnote{It may also happen that the larger fraction of the doubly ionized wavepacket is in $D_1$ but then, typically its main component is not along the $u=0$ line, indicating not a true back-to-back motion.}

Here, our goal is to reverse this situation and to increase the occupancy of region 
$D_1$, compared to region $D_2$. 
The optimal choice of target would be a wavepacket that is doubly ionized \new{with large $v$} but centered along $u=0$.
Since the external laser field couples only with the $u$ direction, this can be 
achieved solely by an interplay between the external field and the Coulomb repulsion of the electrons.

\section{Control methods}
\label{sec:control_methods}
The goal of quantum control theory is to steer \emph{via} an external electromagnetic field (or potential) $A(t)$ the evolution of the wavefunction $\ket{\psi(t)}$. 
The evolution is determined by the time-dependent Schrödinger equation (TDSE)
\begin{align}
  \ii \partd{}{t} \ket{\psi(t)} &=  \hat H \ket{\psi(t)} \label{eq:oct_tdse}\\
  &= \hat H_0 \ket{\psi(t)} + \hat X A(t) \ket{\psi(t)},\quad 
A(t) \in \mathbb C, 
\end{align}
where $\hat H_0$ is the field-free Hamiltonian and $\hat X$ the (typically electromagnetic) interaction operator, which is multiplied with the field $A(t)$.
In control theory, the field is treated as a parameter in order to optimize for a specific goal. 

A wide variety of optimization goals
can be formulated in terms of a control functional $J$ that solely depends on the wavefunction at final time $T$. It takes the form of
\begin{equation}
 J = \matrixe{\psi(T)}{\hat O}{\psi(T)},\label{eq:oct_J}
\end{equation}
which should be maximized.\cite{oct_kosloff_1989,tannor_book,oct_rev_brown_2008,oct_rev_werschnik_2007}
$\hat O$ is Hermitian and either some operator whose expectation value at final time $T$ should be optimized, or, more commonly, a projection operator $\ketbra{\phi}{\phi}$ in order to optimize for a particular state $\ket{\phi}$. 
As explained in \autoref{sec:model_system}, the operator $\hat O$ should be chosen in such a way that maximizing $J$ corresponds to the major contribution of the wavefunction coming from region $D_1$. For example, this can be achieved either by choosing $\hat O$ to be the position operator $\hat O = v$~\footnote{To simplify notation, we drop the hats in the coordinate operators $\hat u$ and $\hat v$. It will be clear from the context whether operators are meant or not.}, or to be a projector, $\hat O=\ketbra{\phi_v}{\phi_v}$, where $\phi_v$ is a state localized in $D_1$.
Other choices will be discussed later.

Quantum control theory can be implemented in various ways.
We start our analysis with local control theory in \autoref{sec:theory_lct} and show the difficulties of using this method for our objective.
While we do not include results of local control theory in the main text, 
\footnote{Local control results are presented in \new{Section C.1} of the supplementary material.}
the analysis will be helpful for understanding more about the interplay of the field with our objective and it will help us choosing appropriate gauges and definitions of objectives for the other control algorithms used. 

Following that, we present the
three different quantum control algorithms that are actually used in the main text, short-time control (\autoref{sec:theory_short_time_control}), derivative-free optimization (\autoref{sec:theory_deriv_free_opt}) and the Krotov method (\autoref{sec:theory_krotov}). Furthermore, we apply optimal control to the classical equations of motions (\autoref{sec:theory_class}). All methods have advantages and disadvantages and may lead to different pulses and hence different physical mechanisms. However, in this work, we show that \emph{all} four control methods (including the classical control method) lead to essentially the same physical mechanism. This is discussed in \autoref{sec:discussion}. Note that since the different control methods involve different numerics, the form and choice of target operator (state or projector) may have to be modified depending on the control methodology.

\subsection{Local Control}
\label{sec:theory_lct}

One way to maximize \autoref{eq:oct_J}, that is, some property of the wavefunction at final propagation time, is to maximize this property \emph{during} the time propagation for all \emph{intermediate} times.
This is the idea of local control theory. 
There, the field $A(t)$ at time $t_0$ is varied \emph{locally} at time $t_0$ 
for a small time-interval 
$t\in[t_0,t_0+\Delta_t]$ in order to optimize the expectation value of some operator $\hat O$.\cite{local_control_rev_tannor_2009,oct_kosloff_1989}
In the following, we will assume that $\hat O$ is time-independent.
According to Ehrenfest's theorem, the rate of $\erw{\hat O}$ then
becomes%
\begin{align}
  \totd{}{t} \erw{\hat O} &= \ii \matrixe{\psi(t)}{[\hat H_0,\hat O]}{\psi(t)} + 
  \ii A(t) \matrixe{\psi(t)}{[\hat X,\hat O]}{\psi(t)},\label{eq:lct_expr_dt}
\end{align}
where $[\hat A,\hat B]$ is the commutator of operators $\hat A$ and $\hat B$.
Provided that $\hat O$ does not commute with the field interaction operator, $\hat X$, the temporal change of $\erw{\hat O}$ can be controlled by an appropriately chosen $A(t)$.

\autoref{eq:lct_expr_dt} can be written as
\begin{equation}
  \totd{\erw{\hat O}}{t} \equiv C(t) =  \mathcal Y(t) +  A(t) \mathcal Z(t).
  \label{eq:lct_expr_dt_s}
\end{equation}
{%
This equation gives us a means of optimizing $C(t)$ by controlling the field, $A(t)$.
As $\mathcal Z(t)$ depends on $[\hat X, \hat O]$,  
the optimizing expression for $A(t)$ depends on the form of %
$\hat O$ and the form of 
$\hat X$, the latter depending on the gauge. 
}

In the length gauge, $\hat X$ \emph{commutes} with operators that are functions of the coordinates such
that no field-dependent expressions are obtained to first order.
Although local control is, in principle, possible \emph{via} a second-order expression {(see Section \new{B.1} in the supplementary material),} %
this turns out to be very unstable numerically and is not pursued further.

Instead, we employ the velocity gauge. 
For reasons that will become clear below, here, we choose  some operator $\hat O$ instead of an projector for optimization.
Assuming that $\hat O$ depends only on $u$ and $v$ (and not derivatives of $u$ or $v$), $\hat O(u,v)$, inserting \Eqs \ref{eq:he_H_s} and \ref{eq:he_B_s_v} into  \autoref{eq:lct_expr_dt} yields
\begin{equation}
\begin{split}
  \totd{\erw{O(u,v)}}{t} =& %
  - \frac{\ii}{4} \bigl[2 \erw{ O_v(u,v) \partial_v } + \erw{ O_{vv}(u,v) } \\
  &+ 2 \erw{ O_u(u,v) \partial_u }  + \erw{ O_{uu}(u,v) } \bigr]\\
  &+A(t) \erw{ O_u(u,v) },
\end{split}
\label{eq:lct_o_dt}
\end{equation}
where $O_{uu}(u,v) = \partdd{O(u,v)}{u} \equiv \partial_u^2 O(u,v)$ and so on.
As noted in \autoref{sec:model_system}, the field couples only with
the
$u$ direction. This is evident from the form of \autoref{eq:lct_o_dt}, where $A(t)$
couples solely with $\erw{ O_u(u,v) }$,
{%
which needs to be non-vanishing for controlling $\erw{\hat O}$.
Hence, product forms such as $\hat O(u,v) = \tilde O(u) \cdot \check O(v)$
or non-separable functional expressions for $\hat O(u,v)$ are required, and there must be a dependence of $\hat O$ on $u$.
}

Besides the necessity to use non-separable forms, in order to steer the wavefunction into region $D_1$ the operator $\hat O(u,v)$ also needs to have reasonable overlap with the origin where the ground state, $\ket{\psi(0)}$, is localized. This is the reason why for the local control procedure $\hat O$ should be an operator and not a projector of a final state located solely in $D_1$.

Comparing \autoref{eq:lct_o_dt} with \autoref{eq:lct_expr_dt_s} reveals that $\mathcal Y(t)$ is nonzero.  
This is in contrast to previous applications of local control theory, where $\hat O$ commuted with $\hat H_0$, resulting in vanishing $\mathcal Y(t)$.\cite{lct_ohtsuki_1998,local_control_rev_tannor_2009}
In that case, \autoref{eq:lct_expr_dt_s} takes a particularly simple form and $\erw{\hat O}$ is maximized by setting $A(t)$ to the complex conjugate of $\mathcal Z(t)$.
Here, this procedure does not work. An attempt to find an expression by enforcing $C(t)$ to have a predetermined functional form is shown in Section \new{B.1} in the supplementary material. However, we observed numerical difficulties with this approach, leading to very jagged pulses that do not maximize $\erw{\hat O}$ at all time steps. 
In principle, control via $\totdd{\erw{O(u,v)}}{t}$ is  possible but the explicit dependence  of $\totdd{\erw{O(u,v)}}{t}$ on $A^2(t)$ and $\totd{A(t)}{t}$ also leads to numerical difficulties.

Due to the numerical difficulties of using local control for our objective, we do not include the results of this procedure in the main text, but in the supplementary material.
\footnote{Local control results are presented in Section \new{C.1} of the supplementary material.}
However, its analysis gained insights into the optimal choice of gauge and the form of $\hat O$.

\subsection{Short-time control}
\label{sec:theory_short_time_control}
As alternative to local control, we use a short-time control optimization. 
Specifically, $\ket{\psi(t+\Delta_t)}$ is evaluated explicitly and the new $A(t+\Delta_t/2)$ is optimized numerically (a one-dimensional maximization problem), without the need for analytical formulae for time-derivatives.
More details on the procedure are presented in Section \new{B.2} in the supplementary material.
Strictly speaking, for a given time step $\Delta t$,
the short-time control algorithm includes higher-order effects. 
As will be apparent below, this leads to a loss of the local control property of a monotonically  increasing objective.
However, in the limit of $\Delta t \to 0$, 
the standard local control expression can be obtained (and with it the associated numerical difficulties, as discussed above).

\subsection{Global Control}
An alternative to local quantum control is \emph{global} optimization procedures
that maximize the final state or an expectation value by varying the parameters of the field
simultaneously at all times~\footnote{Here, global optimization means not an optimization to a global minimum but that the field is optimized \emph{globally}.}.
In contrast to local control, this means that
the target expression need not monotonically increase during all propagation times.\cite{local_control_rev_tannor_2009}
Instead, the target can be chosen to depend only on the state at final propagation time.
This relaxes the constraints that local control puts on the optimization procedure such that other fields
and other physical processes may be found.
Compared to local control methods, global control methods are typically much more resource-intensive as they require an iterative approach where the TDSE is solved many times until the optimal field is found, whereas local control procedures require only one time propagation.

Global control can be implemented in different ways, two of which will be presented in the following.

\subsubsection{Derivative-free optimization}
\label{sec:theory_deriv_free_opt}

A simple scheme for global control optimization is to represent the field in terms of \new{$N_g$} functions $g_i$ and to optimize the (possibly nonlinear) parameters,
\begin{equation}
 A(t) \approx \sum_{i=1}^{N_g} g_i(t; \vec p_i). \label{eq:fbr}
\end{equation}
The advantage (and disadvantage) of this approach is that the form of the field is predetermined by the form of $g_i$. With that, the spectral bandwidth can be limited or a particular pulse shape (e.g., Gaussian for $N_g=1$) can be required. This, however, results in a constrained optimization of $A(t)$, making the optimization procedure often more difficult.

For finding optimal parameters $\vec p_i$, in principle, the gradient of $J$ with respect to the parameters can be derived using the approach from the next \autoref{sec:theory_krotov}. 
However, especially for nonlinear optimizations, the gradient can be noisy. This renders standard gradient-based optimization algorithms useless.
Instead, black-box derivative-free optimization algorithms are typically used.\cite{derivative_free_optimization_book}
These can be either heuristic algorithms that search for global minima of the optimization landscapes (like genetic algorithms) or algorithms for finding local minima by estimating the gradient numerically.
Both types of algorithms will be used in this work.

Due to the more numerical nature of the derivative-free optimization, the objective does not need to be of the particular form shown in \autoref{eq:oct_J}. 
Instead, \emph{any} objective can be chosen. Here, we chose to maximize the overlap with two states:%
\begin{equation}
  J = | \braket{\psi(T)}{\phi_1} - w_2 \braket{\psi(T)}{\phi_2} |,\label{eq:deriv_free_J}
\end{equation}
where $w_2$ is a weight. The states $\ket{\phi_i}$ can take \emph{any} form and do not even need to be square integrable. 
The overlap of $\ket{\psi(T)}$ with $\ket{\phi_1}$ is maximized. Hence, $\ket{\phi_1}$ should be located in region $D_1$  (compare with \autoref{fig:fig_he_regions}).
In contrast, the overlap of $\ket{\psi(T)}$  with $\ket{\phi_2}$ is minimized. Hence, 
$\ket{\phi_2}$ should be located in region $D_2$. 
Furthermore, for the electrons to move into opposite directions, there should be no momentum in $u$ but a positive momentum in $v$, giving an additional property $\ket{\phi_i}$ should fulfill. 

\subsubsection{Krotov method}
\label{sec:theory_krotov}

Another approach to global quantum control takes derivative information into account.
This is performed within a variational framework where $A(t)$ is discretized,
leading to a very robust and general approach. However, adding particular constraints to the field is difficult (see \lits{oct_constraints_gollub_2008,oct_constraints_schroder_2009} for some approaches) and some target operators $\hat O$ may lead to numerical instabilities due to the lack of constraints (i.e., very large values of $A(t)$ or many oscillations). $\hat O$ thus needs to be well-chosen.

To obtain the optimized field in discretized time, we use the Krotov method.\cite{krotov_book,tannor_control_1992,eitan_optimal_2011}
Therein, not only $J$ but the following functional is optimized:
\begin{align}
 \bar J &= J + J_1 + J_2 + J_3,
 \end{align}
 with 
\begin{align}
  J &= \matrixe{\psi(T)}{\hat O}{\psi(T)},\label{eq:J}\\
 J_1 &= - 2 \Re \int_0^T \braket{\chi(t)}{\partial \psi(t)/\partial t} \dd t,\label{eq:J1}\\
 J_2 &= +2 \Re \int_0^T \matrixe{\chi(t)}{[\hat H_0+\hat X A(t)]}{\psi(t)} \dd t,\label{eq:J2}\\
 J_3 &= - \int_0^T \lambda(t) [A(t) - A^\text{ref}(t)]^2 \dd t.\label{eq:J3}
\end{align}
Here, $\bra{\chi(t)}$ is a Lagrange multiplier or dual function.
The functional $J_3$ restrains the fluence of the field.
The smaller the $\lambda(t)$, the smaller the
contribution of $J_3$ to $\bar J$ and hence the larger the allowed deviations from the reference field $A^{\text{ref}}$.

After discretization in $t$, this optimization problem finally yields the following update of the field at iteration $p$ at time $t$:
\begin{equation}
    A^{(p)}(t) = A^\text{ref}(t) + \frac1{\lambda(t)} \Im 
    \Matrixe{\chi^{(p-1)}(t)}{\hat X}{\psi^{(p)}(t-\Delta_t)}.\label{eq:krotov_field_update}
\end{equation}
In the Krotov algorithm, the dual state $\ket{\chi(T)}$ at final propagation time $T$ is 
obtained by applying $\hat O$ on $\ket{\psi(T)}$. The dual state is then propagated backward in time. The new wavefunction $\ket{\psi(t)}$ and the new values of the field are then obtained by a forward propagation using \autoref{eq:krotov_field_update}. This procedure is repeated until convergence.

Provided that $\hat O$ is a positive semidefinite operator, \eg, a projector, it can be  shown that Krotov's method 
gives monotonic convergence with respect to $\bar J$. 
However, it may happen that only $J_3$ in $\bar J$ is optimized and not the actual target $J$.
This can be avoided by setting the reference field
$A^{{\text{ref}}}$ to the field of the previous iteration, $A^{(p-1)}$. Then, the method
is monotonically convergent both with respect to $\bar J$ and
$J$.\cite{eitan_optimal_2011}

The objective in the Krotov optimization can be an expectation value or a particular state (using a projection operator as $\hat O$).
It is not immediately clear which choice will lead to a better result. Numerically, we found that optimizing for an expectation value is much easier.
The major reason is that,
when optimizing for a particular state, the Krotov optimization is very sensitive to the functional form of the state. 
Since the target state will actually be used for the backpropagation, it is very important that the target state is physical and can be reached with a reasonable pulse shape.
This is different from derivative-free optimization where the target state does not need to be well-behaved.

\subsection{Control of classical dynamics}
\label{sec:theory_class}

The previous methods dealt with control algorithms for quantum mechanics.
However, optimal control can also be formulated for classical mechanics.
Performing both classical and quantum control optimizations on the same systems pinpoints the quantum effects and gives new insights. Practically, the classical equations of motions are, of course, much easier to solve.

For controlling classical mechanics,  Hamilton's equations of motion
\begin{align}
  \totd{\vec p}{t} &= - \partd{H}{\vec x}, \quad \totd{\vec x}{t} = \partd{H}{\vec p}
  \label{eq:hamilton_class_eom}
\end{align}
are solved for the classical Hamiltonian $H$.
For control optimizations, a classical Lagrangian can be formulated
similar to the Lagrangian  formulation of the Krotov method (\autoref{sec:theory_krotov}). The equations are then solved on a time grid as in the Krotov method by discretizing the equations of motions using, e.g., the fourth-order Runge-Kutta propagator.\cite{deuflhard_book2}
The control procedure can then be performed using standard black-box software.\cite{casadi}

\section{Results}
\label{sec:results}

We now present the outcome of the various control methods.
\new{To solve the time-dependent Schrödinger equation we use the dynamically pruned discrete variable representation (DP-DVR).\cite{pW_tannor_2016,pvb_rev_tannor_2018,DCO_hartke_2018}
In the DP-DVR, the direct-product DVR basis is pruned in a non-direct-product fashion at each time step. This allows for a very efficient representation of the wavepacket at all propagation times. We found it faster than FFT-based dynamics, even when the latter is performed on a general purpose graphical processing unit (GPGPU). More details are presented in Section A of the supplementary material.
} 

It will be shown later in \autoref{sec:discussion} that, despite the superficial differences of the optimized pulses, they all share common characteristics and they all result in the same physical mechanism, namely a two-step procedure where the wavefunction is first displaced in $u$ and then evolves into region $D_1$.

We further note that we have tested many different operators and show here only those choices that lead to the most satisfying results. This explains why we sometimes use slightly different functional forms of the operator for the different control methods.
However, in general, we found that operators that maximize region $D_1$ in coordinate space are most useful.  Operators depending on the expectation values of the momenta have also been tested but were found to be less useful.
Further details of the simulation procedures are given in {Section \new{B} in the supplementary material.} %
Our results from local control theory are not discussed in the main text but in Section \new{C.1} of the supplementary material.
\new{Spectrograms of all fields shown are given in Section D of the supplementary material.}

\subsection{Short-time control}

For the short-time control procedure, we need to specify the form of 
the operator to be maximized, $\hat O(u,v)$.
For simplicity, 
we chose $\hat O(u,v) = \tilde O(u) \cdot \check O(v)$. 
To confine the region to be optimized in $u$, we used a Gaussian form, 
$ \tilde O(u) = N  \exp( -\alpha u^2 )$, with $N$ being a normalization factor and $\alpha=\unit[2/5]{a.u.}$
To ensure both reasonable overlap around the origin and a steering to larger $v$ values, we chose $\check O(v)=v^2$.

\begin{figure*}
 \includegraphics[width=.75\textwidth]{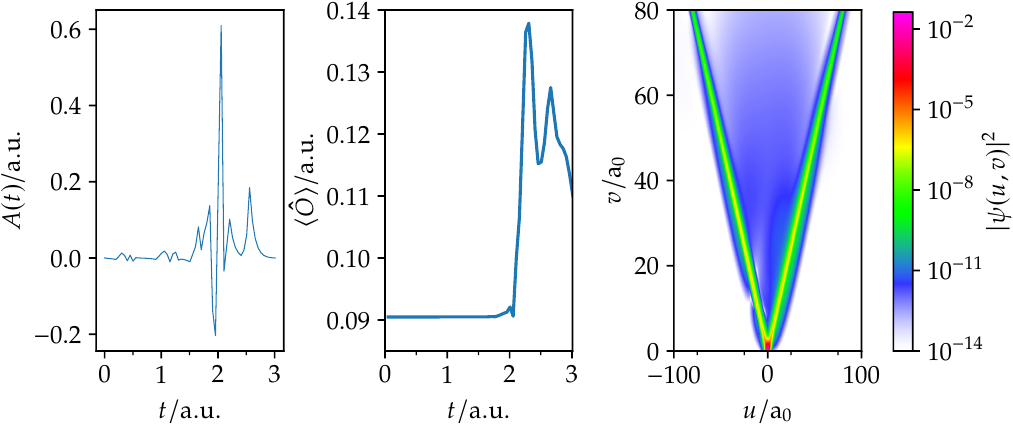}
  \caption{Short-time control optimization.
  The left panel shows the optimized pulse, 
  the middle panel shows the  expectation value of $\mathcal O = N \exp(-2 u^2/5) v^2$ to be optimized,
  and the right panel  shows the square of the wavefunction at the final propagation time $T=\unit[60]{a.u.}$
The short-time control procedure (see \autoref{sec:theory_lct}) is used for the optimization until $t=\unit[3]{a.u.}$ Afterwards, the  wavefunction is propagated field-free.
The colormap has been taken from \lit{brehm_diss}.
  }
  \label{fig:oct_lct_res}
\end{figure*}

The result using the short-time control optimization described in \autoref{sec:theory_short_time_control}  is presented in 
\autoref{fig:oct_lct_res}.
The field obtained is complex and shows many oscillations with jagged features. 
Notably, 
 $\erw{\hat O}$ is not monotonically 
increasing.
This is due to a restriction on the maximal field amplitude\footnote{See  Section \new{B.2} in the supplementary material for details on the restriction on the maximal field amplitude.} %
and the nature of the short-time optimization process as discussed further in \autoref{sec:disc_lct_krotov}.
Even though single ionization dominates, 
there is a clear increase in the objective and there is larger occupancy in 
region $D_1$, compared to region $D_2$. 
The results are discussed further in \autoref{sec:discussion}.

\subsection{Derivative-free optimization}
\label{sec:res_deriv_free}

\new{
In the following, we present three different results %
that differ in 
particular states $\ket{\phi_i}$ used to define the objective $J$ (smooth triangles 
based on the  $\tanh$ function or Gaussians)}:
\begin{align}
  \phi_{1a}(u,v) &= N \exp(-\alpha u^2) \exp[-\alpha (v-v_0)^2 + \ii p_0 (v-v_0)],
 \label{eq:target_gaussian}\\
 \phi_{1b}(u,v) &= 1 - \{ \tanh[ s\cdot  \sin(\pi/4) (a(u) - v + v_0) ] + 1 \} /2,\label{eq:triangle_fu_1}\\
 \phi_{2}(u,v) &= \{ \tanh[ s \cdot \sin(\pi/4)  ( a(u) - v - v_0) ] + 1 \} / 2, \label{eq:triangle_fu_2}
\end{align}
\new{with 
$a(u)$ being a regularized function for taking the absolute value of $u$:}
\begin{align}
 a(u) &= u^2 / \sqrt{u^2 + 0.004}.
\end{align}
\new{The objectives are then defined by}
\begin{align}
 J_{a} &= \braket{\psi(T)}{\phi_{1a}} - 10 \cdot \braket{\psi(T)}{\phi_2},\label{eq:deriv_free_Ja}\\
 J_{b} &= \braket{\psi(T)}{\phi_{1b}} - 1 \cdot \braket{\psi(T)}{\phi_2}.\label{eq:deriv_free_Jb}
\end{align}
Choosing the parameters $\alpha=2/39,\ p_0=3,\ v_0=70$ positions the Gaussian $\phi_{1a}(u,v)$ in region $D_1$.
$\phi_{1b}(u,v)$ is a smooth triangle located in region $D_1$ with $v_0=40$.
In addition, $\phi_2(u,v)$ is a smooth double triangle but it is located in region $D_2$ with $v_0=65$ and $s=0.3$.

For the pulse parametrization, inspired by \lits{ruhman_chirp_1990,cao_chirp_1997,vN_control_brixner_2010},
we chose the real part of a single complex Gaussian
with complex width parameter $\alpha$, which enables chirped fields:
\begin{equation}
g(t) = A_0 \exp[-\Re \alpha  (t-t_0)^2] \cos[(t-t_0) (\omega_0 - \Im \alpha  (t-t_0))].
\label{eq:field_fu_vN}
\end{equation}
It turned out that optimizing one single function is sufficient.
The form of \autoref{eq:field_fu_vN} leads to the following five parameters $\{A_0,\Re \alpha, \Im\alpha, 
\omega_0,t_0\}$ to 
be optimized. 

The choice of objectives and optimization algorithms for the pulses (a)-(c) are as follows.
Pulse (a) used $J_a$ in \autoref{eq:deriv_free_Ja}. 
The other pulses used $J_b$ in \autoref{eq:deriv_free_Jb}.
For pulses (b) and (c), $A_0$ was fixed and not optimized. 
Pulses (a) and (c) were results of local, derivative-free optimizations \emph{via} the COBYLA~\cite{COBYLA} and the BOBYQA~\cite{BOBYQA} algorithms, respectively.
\new{The initial guess pulse was arbitrarily chosen to have the parameters  $\{A_0,\Re \alpha, \Im\alpha, 
\omega_0,t_0\}\sim \{2.3,0.07,-0.17,10.8,5.9\}$, and does not lead to the desired objective.} 
Pulse (b) was a result of global optimization \emph{via} the differential 
evolution algorithm.\cite{weise_book2011}
See Section \new{B.3} in the supplementary material %
for further details. 
We note that the results we present here may not be ``optimal'' compared with other results using the same optimization algorithm.
Instead, we chose those fields that led to a simple pulse shape and had a clear visual appearance of our control aim (major part of the wavefunction in region $D_1$).

\begin{figure*}
  \centering
 \includegraphics[width=.9\textwidth]{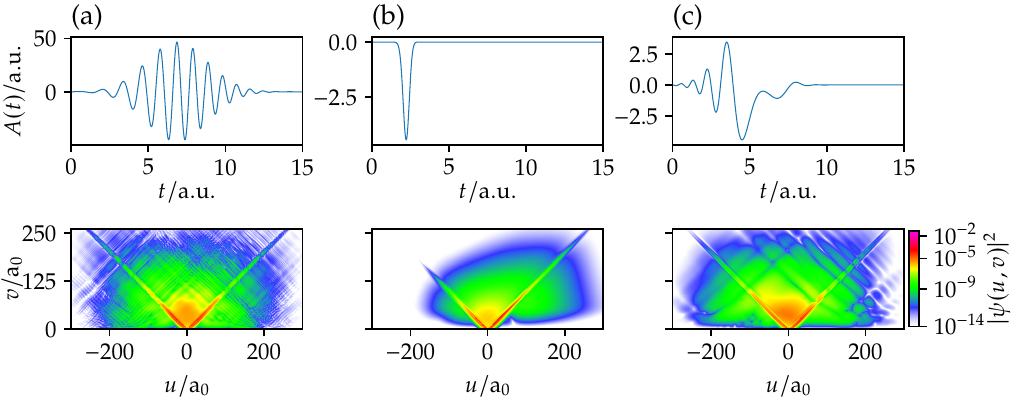}
  \caption{Derivative-free optimization. The pulse with parameters to be optimized is shown in \autoref{eq:field_fu_vN}. 
  The left panels (a) show the COBYLA-optimized pulse (local optimization) using the objectives described at \autoref{eq:target_gaussian} and \autoref{eq:triangle_fu_2}.
  The middle panels (b) show the pulse optimized \emph{via} differential evolution (global optimization), where the objective of \autoref{eq:target_gaussian} is replaced by \autoref{eq:triangle_fu_1}.
  The right panels (c) show the BOBYQA-optimized pulse (local optimization) (same objective as (b)).
  The top panels show the optimized pulses.
  For $t>\unit[15]{a.u.}$, all pulses have vanishing values. See \autoref{tab:oct_DerivFree_res} 
    for the pulse parameters.
  The lower panels show the square of the wavefunction at the final propagation time $T=\unit[60]{a.u.}$
}
  \label{fig:oct_DerivFree_res}
\end{figure*}

The three obtained pulses and the wavefunction at the end of the propagation time are shown in 
\autoref{fig:oct_DerivFree_res}. 
\new{Since all fields vanish after $t=\unit[15]{a.u.}$, we show them only until this time. The final propagation time was $T=\unit[60]{a.u.}$.}
The parameters for \autoref{eq:field_fu_vN} are 
stated in \autoref{tab:oct_DerivFree_res}.
Note that these three pulses are only a selection of a plethora of different pulses, obtained by different algorithms and different targets, but 
their shapes are characteristic of the other pulses as well.

All pulses shown lead to an increase of the wavefunction in region $D_1$, compared to 
region $D_2$, such that they fulfill the requested task. 

\begin{table}[!htbp]
 \caption{Derivative-free optimization. Rounded values of the optimized parameters for the 
pulses depicted in \autoref{fig:oct_DerivFree_res}. The form of the pulse is given by 
\autoref{eq:field_fu_vN}. Atomic units are used unless stated otherwise.} 
 \label{tab:oct_DerivFree_res}
 \begin{ruledtabular}
 \begin{tabular}{cdcddddd}
   pulse & \multicolumn{1}{c}{$A_0$} &  \multicolumn{1}{c}{$I [\unit{W/cm^2}]$} &  \multicolumn{1}{c}{$\Re \alpha$} &  \multicolumn{1}{c}{$\Im \alpha$} &  \multicolumn{1}{c}{$t_0$} & 
 \multicolumn{1}{c}{$\omega_0$} &  \multicolumn{1}{c}{$\omega_0 [\unit{eV} ]$}\\\hline
(a) & 46.6& $8 \cdot 10^{19}$ & 0.13 & -0.15& 7& 5.90 & 160\\
(b) & -4.43& $7\cdot 10^{17}$ & 12 & 0.74& 2& 1.43& 38.8\\
(c) & -4.43 & $7\cdot 10^{17}$ & 0.24& -0.85& 5 & 2.19 & 59.3\\
 \end{tabular}
\end{ruledtabular}
\end{table}

\subsection{Krotov}
\label{sec:res_krotov}

For the Krotov algorithm, we tried both projection operators (optimizing for a particular state) and general operators (optimizing for expectation values) as optimization targets $\hat O$.  Using a projection operator led to no satisfying results. More details are given in Section \new{C.2} in the supplementary material. %

Here, we present only results for optimizing expectation values of some operator $\hat O^{K}$.
We closely follow the procedure used in \autoref{sec:res_deriv_free}
but used the functional forms of $\phi_{i}$ as expectation values.
In particular, we used the form of two smooth triangles, Eq.~\ref{eq:triangle_fu_1} and \ref{eq:triangle_fu_2}, from \autoref{sec:res_deriv_free} with shift $v_0$ set to $40$.  That is, $\hat O^{K}(u,v) = \phi_{1b}(u,v) - \phi_2(u,v)$.

\begin{figure*}
 \includegraphics[width=1.4\columnwidth]{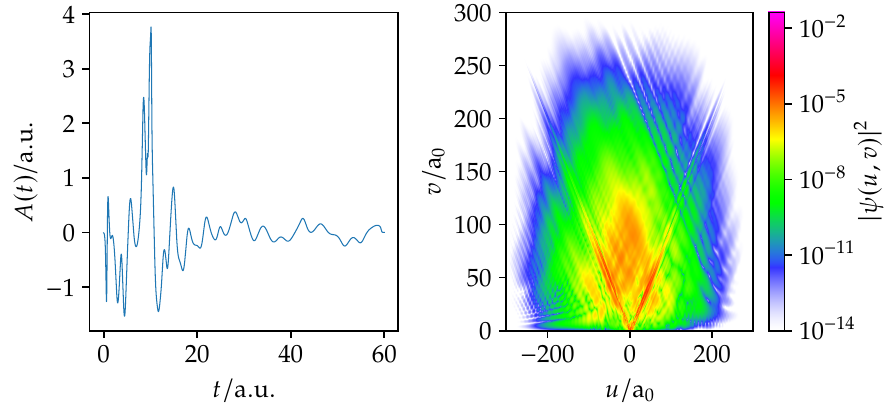}
 \caption{Krotov optimization. The expectation value of Equations \ref{eq:triangle_fu_1} and \ref{eq:triangle_fu_2} is optimized (same as in the derivative-free optimization but with a different shift $v_0=40$; compare with pulses (b) and (c) in  \autoref{fig:oct_DerivFree_res}).
 The left panel shows the optimized pulse.
 The right panel shows the square of the wavefunction at the final propagation time $T=\unit[60]{a.u.}$}
 \label{fig:oct_krotov_expt_val}
 \end{figure*}

\autoref{fig:oct_krotov_expt_val} shows the final field from the Krotov iteration. Clearly, region $D_1$ is more occupied than region $D_2$.

\subsection{Control of classical dynamics}

\begin{figure*}
    \includegraphics[width=.44\textwidth]{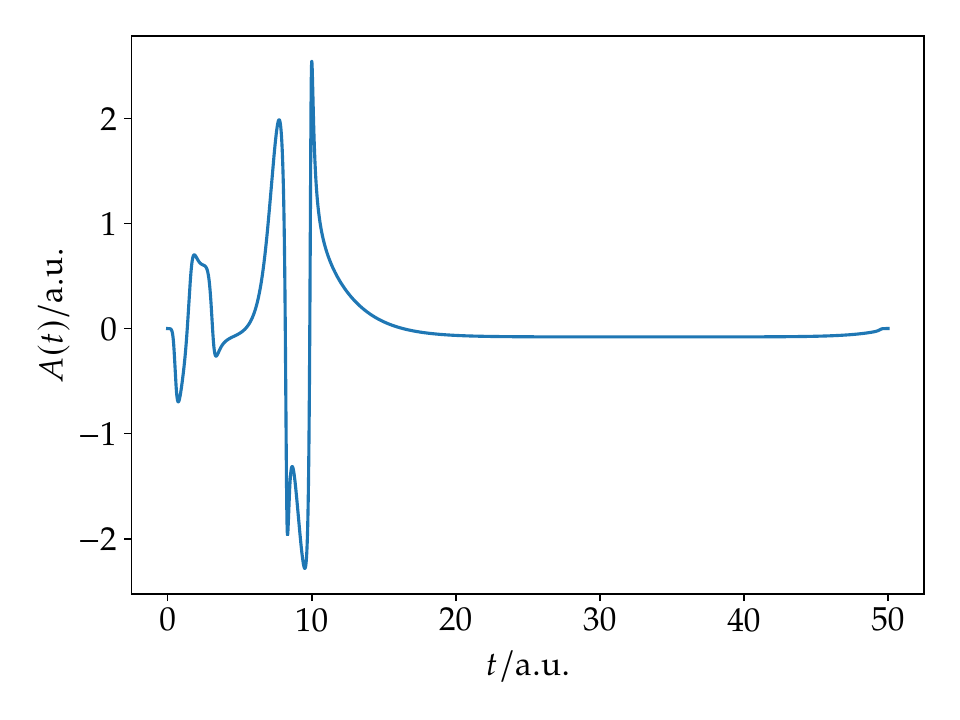}
    \includegraphics[width=.5\textwidth]{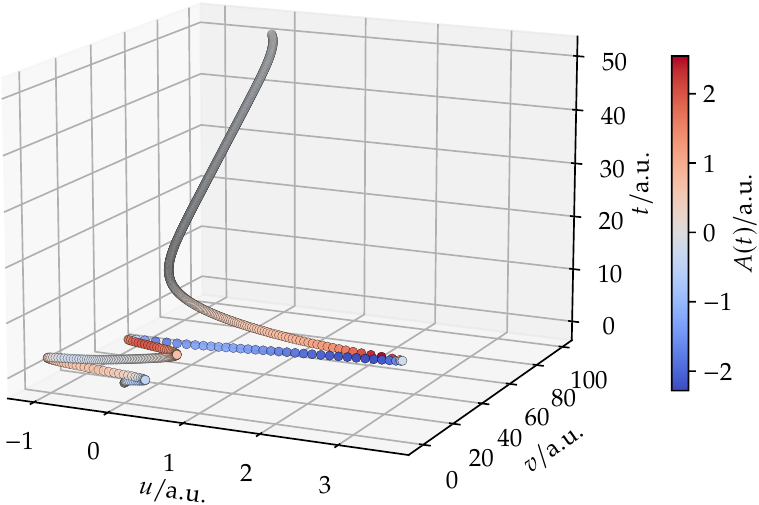}
    \caption{Classical optimal control result.
    The objective is shown in \autoref{eq:classical_control_objective}.
    The left panel shows the optimized pulse.
    The right panel shows the trajectory in $u$ and $v$ as a function of time $t$ ($z$-axis). 
    The color corresponds to the values of the field.}%
  \label{fig:classical_oct}
\end{figure*}

Controlling the classical dynamics turned out to be much easier 
and various objectives gave very similar results.
As classical dynamics is much simpler, even for complicated control problems, we chose a more complicated objective $J_C$ that contains a time-dependent term, namely
\begin{equation}
 J_C = - \eta v(T) p_v(T) + \int \dd t [u(t)\tanh(\xi t)]^2
\label{eq:classical_control_objective}
\end{equation}
with parameters $\eta=100$ and $\xi=0.1$.
The time-dependent term in $J_C$ ensures that large values of $|u|$ occur only during the propagation but not at the end. In $v$, we maximized both the coordinate and the momentum such that the trajectory ends the simulation with an appropriate direction.
The obtained classical trajectory in $(u,v)$ and the optimized field are shown in \autoref{fig:classical_oct}.

\section{Two-Step Mechanism}
\label{sec:discussion}
We will now discuss the wavefunction propagation that takes place for each field and detail the resulting physical mechanism. It will turn out that all pulses share common characteristics and that they all lead to essentially the same physical mechanism.

For a more thorough analysis, both the wavepacket in coordinate space and its reduced densities in
phase space are shown for selected times. For the phase-space analysis, the Husimi Q representation is used, that is, a projection of the state onto coherent states:\cite{schleich_book}
\begin{align}
 Q(x,p) &= \frac{1}{\pi} | \braket{{g_{x,p}}}{\psi} |^2,\label{eq:husimi_distr}\\
 g_{x_0,p_0}(x,p) &= \left(\frac{2\alpha}{\pi}\right)^{\frac{1}{4}}\exp\left[-\alpha(x-x_0)^{2}+\ii\frac{p_0}{\hbar}(x-x_0)\right]
\end{align}
Essentially, the Husimi Q representation is a continuous version of the 
``pixelated'' von Neumann re\-pre\-sen\-ta\-tion,\cite{tannor_vN_beginning_2007,pvb_rev_tannor_2018}
which we have previously used in similar studies.\new{\cite{pvb_math_tannor_2016,pvb_edyn_takemoto_tannor_2012,pvb_edyn_tannor_2015}}
Here, the width parameter $\alpha$ is set to $1/2$, resulting in the same widths in $x$ and $p$.
A phase-space analysis, using Wigner's representation, has also been performed for double ionization by
Lein \emph{et al.}~but only for the $u$ coordinate, focusing on the front-to-back motion of double ionization (same direction {of ionization} for both electrons).\cite{lein_2000}

For the following discussion, recall that large (small) $u$ values
with small (large) $v$ values means that both electrons are on the same (opposite) side
of the nucleus; compare with \autoref{fig:fig_he_regions}.

We start this section with the results from the derivative-free optimization instead of the results from short-time control. The changed order is due to the fact that the outcomes from the derivative-free optimization have the simplest shapes and thus highlight the main characteristics of all the pulses found.
They will also highlight the main physical mechanism that takes place.

\subsection{Derivative-free optimization}
\label{sec:disc_deriv_free}

We first consider the simplest pulse found in our study, namely field (b) obtained from
the derivative-free optimization; see \autoref{fig:oct_DerivFree_res}.
This field consists
of just a single pronounced peak at negative values.\footnote{Note that the field in the length
gauge would have components with both positive and negative values.
Note further that, due to symmetry of the potential in $u$, the overall sign of the field
does not matter.}
The pronounced peak leads to an acceleration of the wavepacket in the $u$ direction.
Since the final wavepacket is localized mainly in regions $S$ and $D_1$ (small $u$ values) but \emph{not} in $D_2$ (large $u$ values), this pulse
is, at first sight, counterintuitive. Nevertheless, an analysis of the wavepacket at
different propagation time reveals a clear mechanism.

\begin{figure*}

    \includegraphics[width=.70\textwidth]{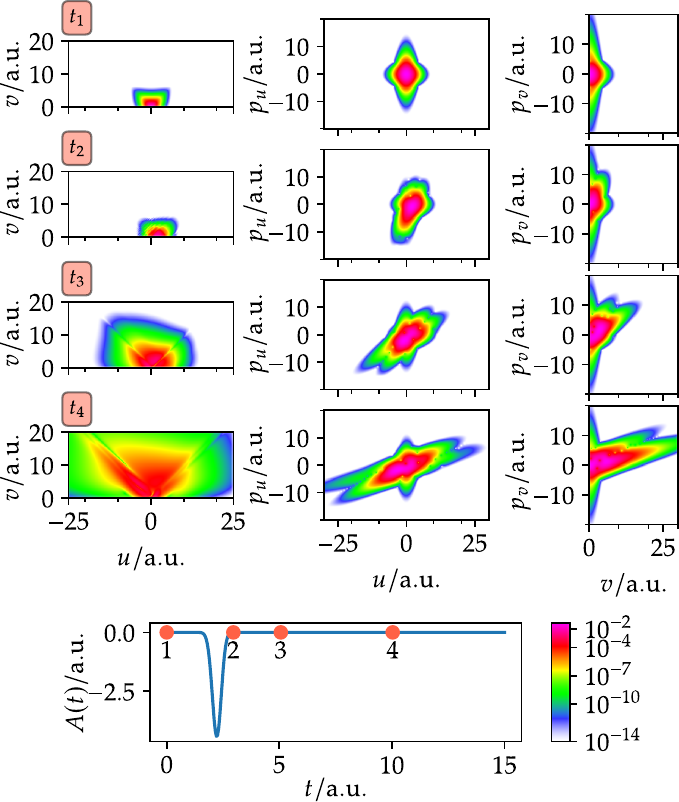}
\caption{%
  Analysis of the mechanism of pulse (b) of the derivative-free optimization (\autoref{fig:oct_DerivFree_res}).
  The upper left panels show the square of the wavefunction in coordinate space $(u,v)$.
  The upper middle and right panels show the Husimi Q distribution (\autoref{eq:husimi_distr}) of the reduced densities (wavefunction integrated over the other coordinate) in $(u,p_u)$ (middle panels)  or $(v,p_v)$ (right panels), respectively. 
Each row corresponds to a different time $t_i$.
The lower left panel shows the pulse and the chosen time points $t_i$. 
}
  \label{fig:oct_mech_field2}
\end{figure*}

\autoref{fig:oct_mech_field2} shows the reduced density of the wavepacket in coordinate space and
phase-space representations for selected propagation times.
(Plotting the wavepacket in momentum space is not useful because it does
not reveal any coordinate-dependent information.)
The pulse generates an acceleration of the wavepacket in the $u$ direction which {in turn} leads to a displacement of
the wavepacket in $u$; compare $t_1$ and $t_2$ (first and second row) in 
\autoref{fig:oct_mech_field2}. 
Some parts of the wavepacket are scattered at the potential valley but the main part of 
the 
wavepacket is simply displaced. The acceleration in $u$ is visible in the Husimi Q 
distribution of that coordinate. 
Note that the maximum of the wavepacket is still at $v\sim 0$ where the electrons are 
affected strongly by their Coulomb repulsion. Therefore, once the wavepacket is displaced 
and the field vanishes, the wavepacket is driven to larger $v$ values to avoid the 
Coulomb repulsion. Additionally, it is attracted by the nucleus and thus driven to 
smaller $u$ values.
The net result of the motion is
diagonal and in the direction of region $D_1$. This is evident from $t_3$ (third row
in \autoref{fig:oct_mech_field2}). 
At that time, the wavepacket is again centered around 
$u\sim 0$ with large components at larger $v$ values. Interference phenomena due to the 
scattering by the potential valley (along the diagonal) are visible. Note how the 
wavepacket 
has gained large positive components in momentum $p_v$ for larger $v$ values; as is 
evident from the Husimi Q distribution 
in 
coordinate $v$.\footnote{The larger momentum components for $v=0$ are due to the form of 
  the potential and as such are ``ground state quantities'' and not very interesting for this study.}
The components in $p_v$ are large enough to drive the wavepacket into the continuum along 
the $v$ direction.
The nuclear attraction (the potential valley) hinders the wavepacket from moving to large 
negative $u$ values.
The phase-space distribution in $u$ becomes almost point symmetric.
Thus, the total movement is toward large $v$ values; see the last row in \autoref{fig:oct_mech_field2}.

\begin{figure*}
    \includegraphics[width=.75\textwidth]{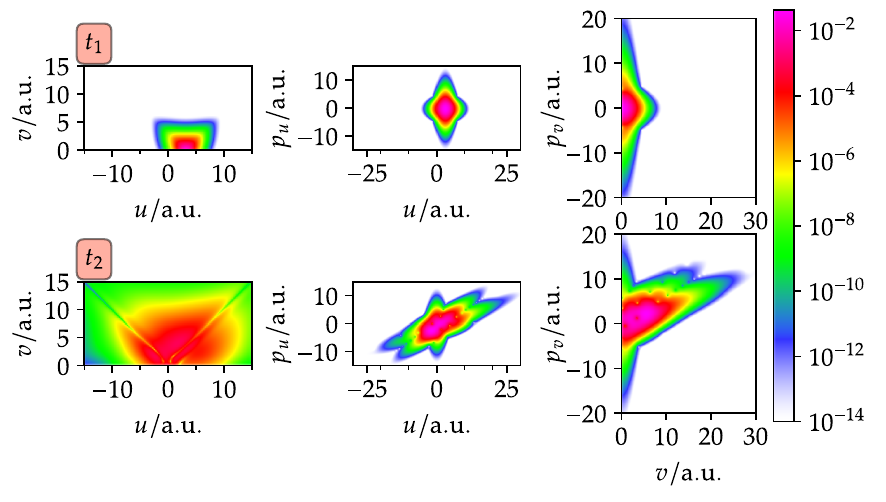}
  \caption{Same as 
\autoref{fig:oct_mech_field2} but without a field and with an initial 
wavepacket displaced in $u$ and centered at $u=\unit[3]{a.u.}$ The propagation times are 
$t_1=0$ and $t_2=\unit[5.02]{a.u.}$
}
  \label{fig:oct_mech_displ}
\end{figure*}

\begin{figure}
 \includegraphics[width=\columnwidth]{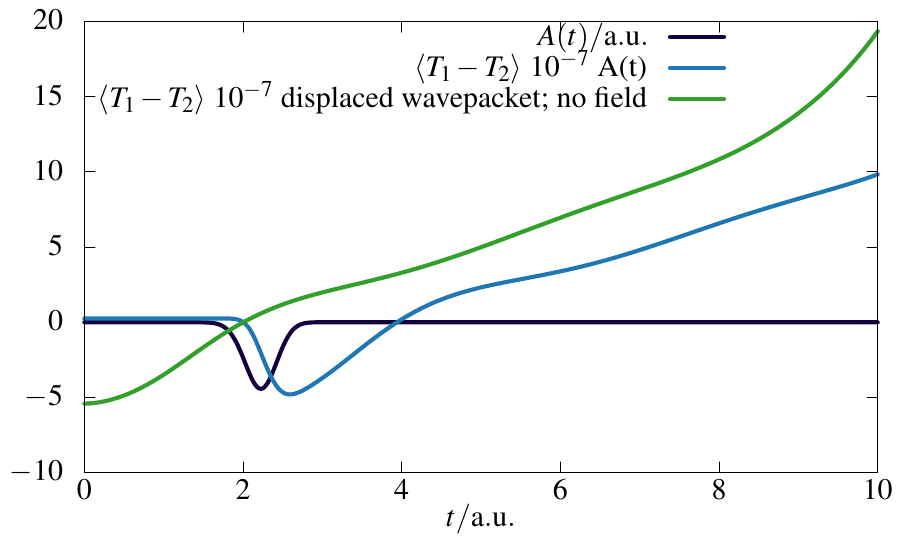}
 \caption{Field (black curve) and optimized expectation values of pulse (b) of the derivative-free optimization (\autoref{fig:oct_DerivFree_res}).
  The blue (green) curve corresponds to the dynamics presented in \autoref{fig:oct_mech_field2} (\autoref{fig:oct_mech_displ}). 
 The observable is $T_1-T_2\equiv \phi_1(u,v) - \phi_2(u,v)$, as defined in \autoref{eq:triangle_fu_1} and \autoref{eq:triangle_fu_2}.}
 \label{fig:expvals}
\end{figure}

To show that the displacement of the wavepacket in $u$ is the first step towards the desired
occupancy in region $D_1$, the propagation of an initially displaced ground state without 
the appearance of an external field is shown in \autoref{fig:oct_mech_displ}. As 
expected, the occupancy in region $D_1$ is visible after a propagation for a short time 
($t_2$; lower row). At that time, there are also major components in 
region $D_2$ but the strong asymmetry of the phase-space distribution in $v$ (lower right 
panel) will drive the wavepacket to larger $v$ values such that the occupancy in $D_2$ 
will vanish.
{This is evident from inspections of the wavepacket at longer propagation times (not shown)
  and from the evaluation in \autoref{sec:common_evaluation}. }
A plot of the field corresponding to pulse (b) and the optimized expectation value of the observable as a function of time, both for the propagation with an external field and for the displaced wavepacket without a field, is shown in \autoref{fig:expvals}.
One clearly sees that the displacement first leads to a decrease in the observable and only afterwards is the observable monotonically increasing.

\begin{figure}
    \includegraphics[width=\columnwidth]{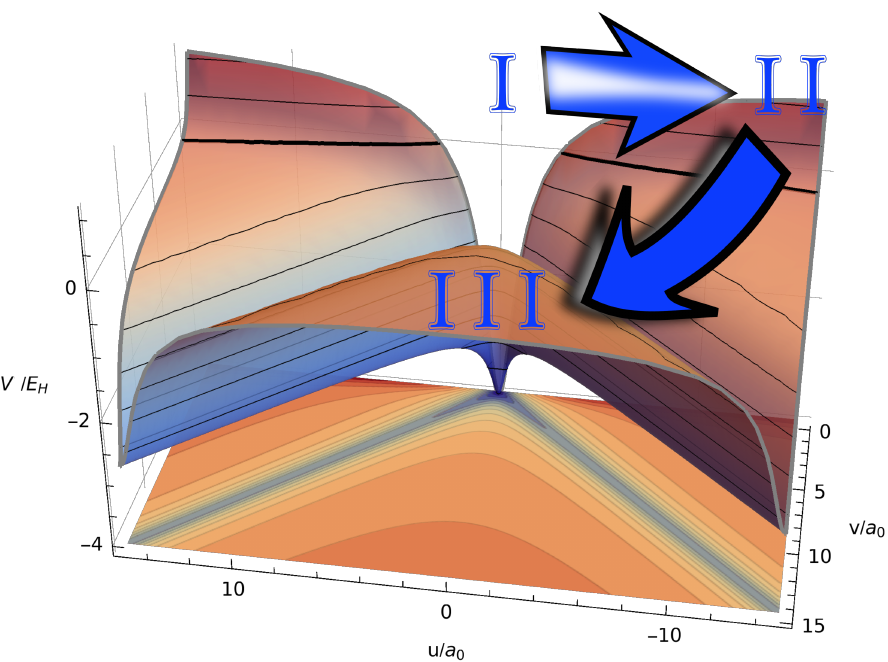}
  \caption{Schematic of the mechanism of the ionization dynamics 
    analyzed in Figs.~\ref{fig:oct_mech_field2}, \ref{fig:oct_mech_displ} and \ref{fig:expvals}.
    Shown are the potential, the initial (I), intermediate (II) and final (III) locations of the wavepacket. The two-step procedure is 
shown by the arrows that represent the movement of the wavepacket.
Note that the potential is symmetric in $v$ and only non-negative values of $v$ are 
shown.}
  \label{fig:symCoulPot_mech}
\end{figure}

Overall, there is a simple two-step procedure that leads to the increased occupancy 
in region $D_1$.
Pictorially, this two-step 
procedure is summarized in \autoref{fig:symCoulPot_mech}:
The wavepacket starts at the origin (position I) and the large values of 
the pulse in one direction leads to a displacement in $u$ in the first step. Once the 
field vanishes, the wavepacket is at position II and the repulsion by the 
electron-electron Coulomb potential and the attraction by the nuclear Coulomb potential 
leads, in the second step, to a movement to region $D_1$ (position III). 

This two-step procedure requires that the field has a strong enough amplitude in order to
displace the electrons in the $u$ direction for small $v$ values.
For these coordinate values, the Coulomb repulsion of the electrons approaches the maximum (see \autoref{fig:symCoulPot}).
This requires that the field interaction has to be stronger than the Coulomb repulsion of the electrons. This requirement is similar to that found in textbook discussions of strong-field
single ionization, where the magnitude of the field is compared to the
Coulomb attraction of the nucleus.

Further tests were performed by scanning the amplitude of the field. An increase in region 
$D_1$ is already visible for a field with amplitude of $\unit[1]{a.u.}$ ($\unit[3.5\cdot 10^{16}]{W\,cm^{-2}}$). 
For lower intensities, only single ionization occurs. For 
amplitudes  larger than $\unit[10]{a.u.}$ ($\unit[3.5\cdot 10^{18}]{W\,cm^{-2}}$), the 
displacement is so large that the wavefunction gains enough momentum in $u$
to overcome the nuclear attraction by the Coulomb potential. This leads to an increase in 
the occupancy in region $D_2$.

\begin{figure*}
    \includegraphics[width=.70\textwidth]{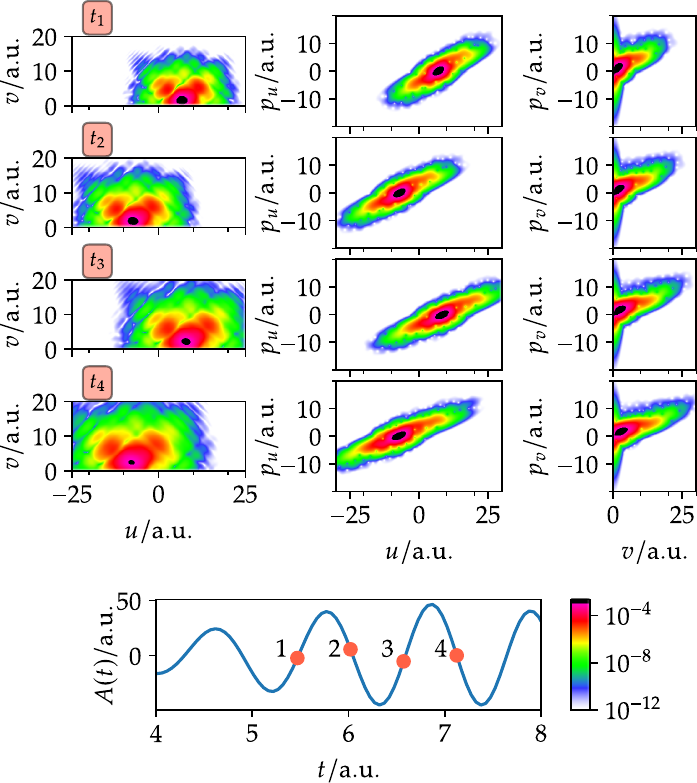}
\caption{Same as \autoref{fig:oct_mech_field2} but with pulse (a) shown in 
\autoref{fig:oct_DerivFree_res} (derivative-free optimization).
The shown times roughly correspond to the zero-crossings of the 7th {to} 10th half-cycles. 
}
  \label{fig:oct_mech_field1}
\end{figure*}

Although the other fields shown in \autoref{fig:oct_DerivFree_res} look quite different,
they actually all follow the same mechanism. 
For example, field (c) has a large time-asymmetry which implies a similar mechanism; an explicit  analysis (not shown here) confirms this.
For pulse (a), 
this is not very clear because there is no evident asymmetry. Indeed, it
is the only pulse that has many, ``symmetric'' cycles (as more standard pulses) but an extremely large amplitude of 
almost $\unit[8.8\cdot 10^{19}]{W\, cm^{-2}}$.  %
 {Reducing the field intensity by a factor of four gives a qualitatively similar result. }
The wavefunctions at different times for 
that pulse are shown in \autoref{fig:oct_mech_field1}. The times roughly correspond to 
different zero-crossings of adjacent half-cycles. At these times, the wavepacket has the 
maximal displacement. Comparing the wavepacket for subsequent times shows that the center 
of the wavepacket is moving to larger and larger $v$ values. This is also evident from
the Husimi Q distribution in $v$. The amplitude of the field is large enough that the wavepacket in the $u$
direction is driven mostly by the continuously oscillating field; as a result
the Husimi Q distribution in $u$ is almost point symmetric (at a displaced point).
Once the field vanishes, the wavepacket has enough positive momentum in $v$
such that the dominant occupancy already created in region $D_1$ remains.

\begin{figure}
    \includegraphics[width=.9\columnwidth]{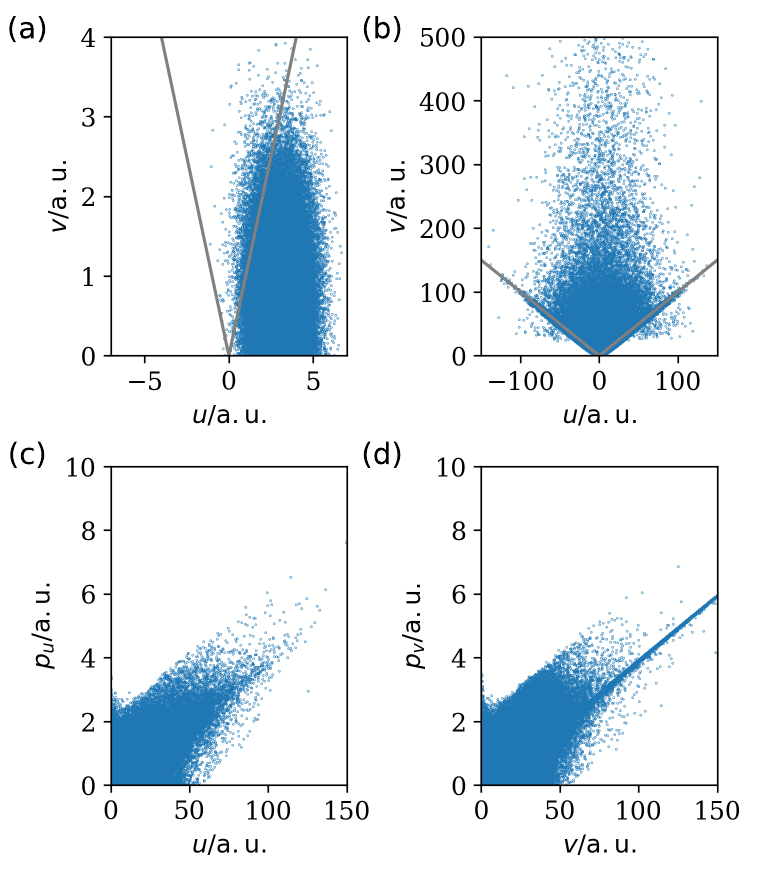}
      \caption{Quasiclassical trajectories corresponding to the quantum mechanical analogue shown in \autoref{fig:oct_mech_displ} (displaced groundstate with field-free propagation).
          Shown are the initial \new{(panels (a) and (c))} and final ($t=\unit[50]{a.u.}$\new{, panels (b) and (d)}) distributions in coordinate \new{(upper panels)} and phase space (lower panels). 
          The gray lines along $|u|=v$ in the upper panels should serve as a visual guide.
      The propagation was performed field-free but the initial starting points were displaced in $u$ by $\unit[3]{a.u.}$}
  \label{fig:qct_dist}
\end{figure}

The two-step procedure is, essentially, a classical mechanism. Indeed, we are able to confirm
this by running quasi-classical trajectories.
The starting conditions of the trajectories were obtained by sampling the four-dimensional phase-space Husimi distribution of the quantum mechanical ground state,  neglecting zero point energy.\cite{bowman_1989,miller_1989}
\autoref{fig:qct_dist} shows the propagation of an ensemble of $10^5$ particles with initial displacement in $u$ of $\unit[3]{a.u.}$
This is the classical counterpart of the quantum mechanical dynamics shown in \autoref{fig:oct_mech_displ}.
Clearly, the final classical phase-space distribution is very similar to that shown in \autoref{fig:oct_mech_displ}: there is a significant portion of particles in region $D_1$ and almost no particles in region $D_2$.
Also the propagation with field (b) (see \autoref{fig:oct_mech_field2}) leads to a very similar final distribution (not shown here).

\subsection{Short-time control and Krotov optimization}
\label{sec:disc_lct_krotov}

Compared to the parametrized pulses obtained from derivative-free optimizations,
the pulses obtained from local control {(\autoref{fig:oct_lct_res})} and from the Krotov optimization {(\autoref{fig:oct_krotov_expt_val})} are much more complex,
making the analysis also more difficult.
However, a phase-space analysis reveals that the fields actually by and large lead to the same mechanism.
{Both fields} 
have a pronounced asymmetry and the mechanism corresponds essentially to the previously detailed two-step procedure.

The short-time control pulse deserves more attention.
In the limit of $\Delta t \to 0$, the short-time control procedure is the same as the local control procedure.
By definition, the local control procedure maximizes the objective at all time steps.
This contradicts the two-step mechanism, where the displacement of the wavepacket in $u$ leads to a decrease in the objective. 
In other words, the local control procedure should not find this mechanism. Here, however, the short-time control procedure with $\Delta t > 0$ differs from local-control as %
the pulse is \emph{numerically} optimized at each time step. If there is no possible increase in the objective at a given time (subject to a restriction on the maximal field amplitude), the procedure thus finds a field with the smallest possible \emph{decrease} of the objective.
With that, the two-step mechanism is possible for the short-time control procedure. 
Nevertheless, the dominant ionization process of the short-time control pulse is single ionization, indicating that local and short-time control is not an adequate method for our objective. The analysis of local control theory in \autoref{sec:theory_lct}, however, helps in designing appropriate control targets and choosing appropriate gauges.

\subsection{Classical control}
The field obtained from classical control also follows, essentially, the two-step procedure. 
The dynamics was already shown  in \autoref{fig:classical_oct}. 
The pronounced negative peak at about $t=\unit[9]{a.u.}$ displaces the classical particles to larger $u$ 
values. Afterward, the particles are driven back to small $u$ and large $v$ values (region $D_1$).
The positive peak for $t>\unit[10]{a.u.}$ increases the displacement in this direction.

\begin{figure}
\includegraphics[width=.9\columnwidth]{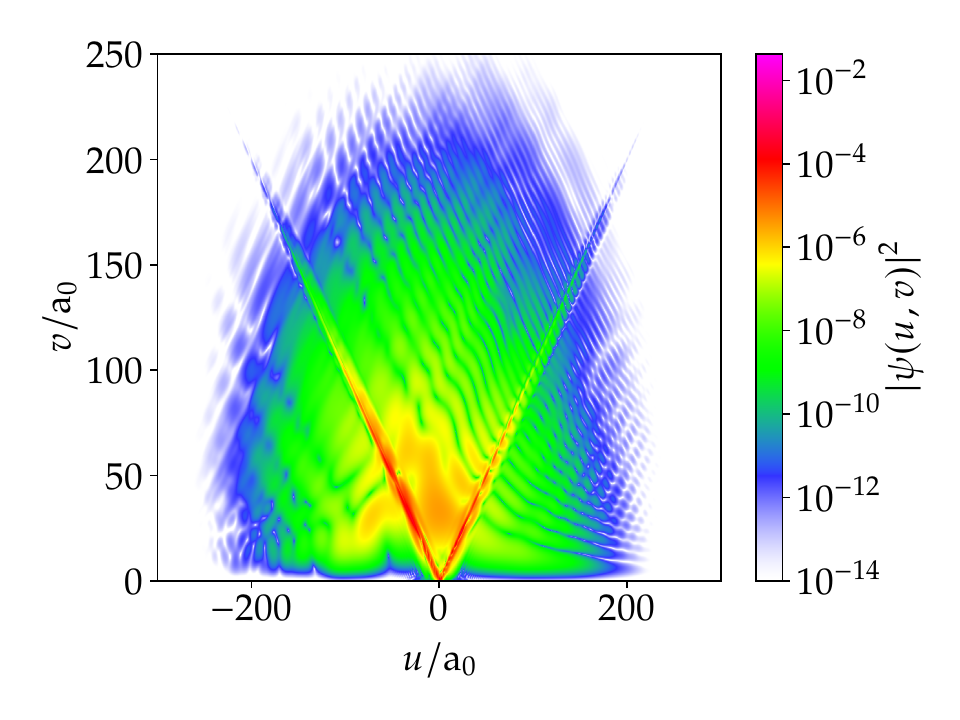}
\caption{Quantum wavefunction propagation with field obtained from classical control
(compare with \autoref{fig:classical_oct}). The square of the wavefunction at the final propagation time, $T=\unit[60]{a.u.}$, is shown.}
\label{fig:prop_field_classical_oct_qm}
\end{figure}

A quantum propagation with this field  is shown in \autoref{fig:prop_field_classical_oct_qm}.
Clearly, an occupancy in region $D_1$ is also obtained here, showing that it is possible
in this case to perform solely classical calculations to obtain qualitatively the same
mechanism.

\subsection{Common evaluation}
\label{sec:common_evaluation}

\autoref{tab:eval} provides a common
evaluation of all results with respect to the occupancies in regions $D_1$ and $D_2$ at the final propagation time.
The occupancies are determined by projectors     $P^{D_1}$ and $P^{D_2}$ that project into the two regions of double ionization shown in  \autoref{fig:fig_he_regions}.
The Krotov method exhibits the maximal expectation value {$\erw{ \hat P^{D_1} - \hat P^{D_2}}$} of $0.283$.
Field (c) from the derivative-free optimization procedure gives the second largest value ($37\,\%$, relative to the Krotov result). This is remarkable as this field is much simpler and less intense than that of the Krotov results (compare \autoref{fig:oct_DerivFree_res} with \autoref{fig:oct_krotov_expt_val}).

Further, for the pulses from the derivative-free optimization procedures, the ratio $\erw{ \hat P^{D_1}} / \erw{\hat P^{D_2}}$ is around $4$ to $6$ times larger than that of the Krotov result. That is, the double ionization yield is higher for the Krotov outcome but the ratio of back-to-back motion versus front-to-back motion is higher for the derivative-free optimization outcomes. 
This is even more extreme for the pulse obtained by the short-time control procedure where double ionization is minimal, but occurs with a strong favor {towards}  back-to-back motion, as is also evident from \autoref{fig:oct_lct_res}.

The highest ratio $\erw{ \hat P^{D_1}} / \erw{\hat P^{D_2}}$ of $403$ actually occurs when the ground state is just displaced and evolved field-free afterwards.
This shows again that, {in our setup}, the displacement in $u$ is an essential process that needs to happen for obtaining back-to-back ionization. 

\begin{table}
\caption{Evaluation of the control results by the 
projectors     $P^{D_1}$ and $P^{D_2}$ that project into the two regions of double ionization shown in  \autoref{fig:fig_he_regions}.
Shown are the expectation values of the difference of the projectors
$P^{D_1}$ and $P^{D_2}$, and the ratio of their expectation values  for the different fields and control methods.
The former (latter) is (in-)dependent on the ionization yield.
The projectors are defined similar to the functions $\phi_{1b}(u,v)$ and $\phi_2(u,v)$ from \autoref{eq:triangle_fu_1} and \autoref{eq:triangle_fu_2}, replacing tanh by step functions.
The expectation values have been calculated at the final propagation time, $T=\unit[60]{a.u.}$
The range of $\erw{ \hat P^{D_1} - \hat P^{D_2}}$ is $[-1,1]$.}
\label{tab:eval}
\begin{ruledtabular}
\begin{tabular}{lllld}
Method & Fig. &  Field &{$\erw{ \hat P^{D_1} - \hat P^{D_2}}$} %
&\multicolumn{1}{c}{$\erw{ \hat P^{D_1}}/ \erw{ \hat P^{D_2}}$}\\ \hline
Short-time Control   &     \ref{fig:oct_lct_res}                               &   &0.000 000 030 & 44\\
Deriv.-Free   & \ref{fig:oct_DerivFree_res} & (a) & 0.063  & 42\\
&                             & (b) & 0.011 & 65\\
&                             & (c) &0.10 & 63\\
Krotov        & \ref{fig:oct_krotov_expt_val} &   & 0.28& 9.9\\
Classical     & \ref{fig:classical_oct}, \ref{fig:prop_field_classical_oct_qm} &    & 0.015& 2.6\\
Displaced     & \ref{fig:oct_mech_displ} &    & 0.035  & 403\\
\end{tabular}
\end{ruledtabular}

\end{table}

{%
We now discuss the validity of our simulation method, namely the nonrelativistic Schrödinger equation within the dipole approximation. 
Relativistic effects have been shown to be important for field intensities around $\unit[10^{20}]{W/cm^2}$.\cite{kjellsson_relativistic_2017}
While field (a) obtained from the derivative-free optimization procedure is in this regime,  further tests  showed that the mechanism is still valid for smaller intensities up to $\unit[3.5\cdot 10^{18}]{W/cm^2}$ (see \autoref{sec:disc_deriv_free}).
All other fields have intensities below $\unit[10^{18}]{W/cm^2}$, where relativistic effects can be neglected. 
We also tested the validity of the dipole approximation by  performing additional simulations that included first and second order corrections to the dipole approximation. 
While we experienced numerical difficulties in simulating field (a), 
all other fields, including the Krotov-optimized, showed only minor changes in the wavefunction when simulated beyond the dipole approximation.
To summarize, the mechanism and the main characteristics of the fields are neither affected by relativistic effects nor are they affected by the dipole approximation. 
}

\section{Conclusions}
\label{sec:conclusions}
We have investigated the control of double ionization dynamics in a (1+1)-dimensional model of the helium atom {for field propagation times up to \unit[1.45]{fs}}. 
{At the onset of ionization,} to first order {in field strength}, an external field can simultaneously accelerate both electrons only into the same direction into the continuum, leading to front-to-back motion of double ionization.
Here, the control objective was to obtain a field that does the opposite, namely back-to-back double ionization, where both electrons are ionized into opposite directions. 
{For the propagation times discussed, }
this can be achieved solely by an interplay between the external field, the electronic repulsion and the nuclear attraction.

We tested four different control procedures.
Three of them are based on quantum mechanics whereas the last one uses classical equations of motions: (1) short-time  control, where the control objective is maximized locally at each time step; (2) a basis expansion of the field and subsequent derivative-free optimization; (3) the Krotov algorithm, where the field is represented on a time grid and derivative information is taken into account; and (4) an algorithm based on classical equations of motions and a collocation-based technique that takes derivative information into account.

Furthermore, we analyzed several variations on the local control method and discussed
why it is not an appropriate algorithm for our objective. 
The local control optimization is not appropriate here because
(1) the operator whose
expectation value is to be optimized does not commute with the field-free
Hamiltonian (i.e.~the expectation value continues to change even after the field is turned off); and (2) %
simultaneous back-to-back motion with the field couples only at second order in time.
Nevertheless, the analysis of the local control method led to insights in choosing an appropriate gauge and the form of the objective.
%

Although the various control algorithms gave superficially different optimal fields, on closer analysis we were able to identify a similar mechanism underlying all the optimal fields based on a two-step procedure. The mechanism is as follows: (1) the electrons are first pushed into the same direction by the field, while they are still close to each other, i.e. front-to-back motion; (2) after the field turns off, the electrons are simultaneously attracted by the nucleus and repel each other. This finally leads to the desired back-to-back motion.
To enable step (1), the potential induced by the field must be stronger than both the Coulomb repulsion of the electrons and the nuclear Coulomb attraction.
Both steps can be repeated several times.
\new{The mechanism displays a nontrivial interplay between the electron-field interaction at a certain intensity range, which drives the electrons to the direction opposite to our goal, and the electron-nuclear and electron-electron interactions, which are both required for steering the displaced electrons to the final back-to-back motion.}
Remarkably, we obtained the same results using classical propagation and optimization. The essentials of this mechanisms are thus a classical effect.

All results have been performed with our new dynamically pruned DVR (DP-DVR),\cite{pW_tannor_2016,pvb_rev_tannor_2018,DCO_hartke_2018}
where a non-direct-product basis is employed that is adapted to the shape of the wavepacket in coordinate space at each time step. Compared to standard, e.g., FFT-based dynamics on a direct-product grid, this approach can be orders of magnitude faster and thus enabled the very quick simulations that were needed to perform this study.

\begin{acknowledgments}
We are thankful to Daniel Reich for helpful discussion about quantum control algorithms and 
Norio Takemoto for helpful discussions in the beginning of this project.
H.~R.~L.~is thankful to Timm Faulwasser and Karl Worthmann for introducing him to classical optimization problems and algorithms during a summer school of the Studienstiftung des deutschen Volkes.
H.~R.~L.~acknowledges financial support from the Fonds der Chemischen Industrie and the Studienstiftung des deutschen Volkes.
Financial support from the Israel Science Foundation (1094/16) and the German-Israeli Foundation for Scientific Research and Development (GIF) are gratefully acknowledged.
\end{acknowledgments}
\new{
\section*{Conflict of interest}
The authors have no conflicts to disclose.
\section*{Data availability statement}
The data that support the findings of this study are available from the corresponding author upon reasonable request.
\section*{Supplementary material}
See supplementary material for more numerical details on our DP-DVR simulation method (Section A) and on the various control algorithms (Section B), for additional control results (Section C), and for spectrograms of the optimized pulses presented here (Section D).
}

\end{document}